\documentclass[pre,twocolumn,showkeys,showpacs,groupaddress,preprintnumbers,floatfix]{revtex4-1}

\usepackage{dynlearn}

\begin{document}

\title{The Origins of Computational Mechanics:\\
A Brief Intellectual History and Several Clarifications}

\author{James P. Crutchfield}
\email{chaos@ucdavis.edu}
\affiliation{Complexity Sciences Center and Department of Physics,\\
University of California at Davis, One Shields Avenue, Davis, CA 95616}

\date{\today}
\bibliographystyle{unsrt}

\begin{abstract}
The principle goal of computational mechanics is to define pattern and
structure so that the organization of complex systems can be detected and
quantified. Computational mechanics developed from efforts in the 1970s and
early 1980s to identify strange attractors as the mechanism driving weak fluid
turbulence via the method of reconstructing attractor geometry from measurement
time series and in the mid-1980s to estimate equations of motion directly from
complex time series. In providing a mathematical and operational definition of
structure it addressed weaknesses of these early approaches to discovering
patterns in natural systems.

Since then, computational mechanics has led to a range of results from
theoretical physics and nonlinear mathematics to diverse applications. The
former include closed-form analysis of finite- and infinite-state Markov and
non-Markov stochastic processes that are ergodic or nonergodic and their
measures of information and intrinsic computation. The applications range from
complex materials and deterministic chaos and intelligence in Maxwellian demons
to quantum compression of classical processes and the evolution of computation
and language.

This brief review clarifies several misunderstandings and addresses concerns
recently raised regarding early works in the field (1980s). We show
that misguided evaluations of the contributions of computational mechanics are
groundless and stem from a lack of familiarity with its basic goals and from a
failure to consider its historical context. For all practical purposes, its
modern methods and results largely supersede the early works. This not
only renders recent criticism moot and shows the solid ground on which
computational mechanics stands but, most importantly, shows the significant
progress achieved over three decades and points to the many intriguing and
outstanding challenges in understanding the computational nature of complex
dynamic systems.

\end{abstract}


\preprint{Santa Fe Institute Working Paper 17-10-XXX}
\preprint{arxiv.org:1710.XXXXX [cond-mat.stat-mech]}

\maketitle

\setstretch{1.1}

\setlength{\parskip}{5pt}
\setlength{\parindent}{0pt}

\section{Goals}
\vspace{-0.1in}

The rise of dynamical systems theory and the maturation of the statistical
physics of critical phenomena in the 1960s and 1970s led to a new optimism that
complicated and unpredictable phenomena in the natural world were, in fact,
governed by simple, but nonlinearly interacting systems. Moreover, new
mathematical concepts and increasingly powerful computers provided an entr\'ee
to understanding how such phenomena emerged over time and space. The
overarching lesson was that intricate structures in a system's state space
amplify microscopic uncertainties, guiding and eventually attenuating them to
form complex spatiotemporal patterns. In short order, though, this new
perspective on complex systems raised the question of how to quantify their
unpredictability and organization.

By themselves, qualitative dynamics and statistical mechanics were mute to this
challenge. The first hints at addressing it lay in Kolmogorov's (and
contemporaries')
introduction of computation theory \cite{Turi36,Shan56c,Mins67} and Shannon's
information theory \cite{Shan48a} into continuum-state dynamical systems
\cite{Kolm56a,Kolm65,Kolm83,Kolm58,Kolm59,Sina59,Chai66}. This demonstrated
that information had an essential role to play in physical theories of complex
phenomena---a role as important as energy, but complementary. Specifically, it
led to a new algorithmic foundation to randomness generated by physical
systems---behavior that cannot be compressed is random---and so a bona fide
measure of unpredictability of complex systems was established.

Generating information, though, is only one aspect of complex systems. How do
they store and process that information? Practically, the introduction of
information and algorithmic concepts side-stepped questions about how the
internal mechanisms of complex systems are structured and organized.
Delineating their informational architecture was not addressed, for good
reason. The task is subtle.

Even if we know their governing mechanisms, complex systems (worth the label)
generate patterns over long temporal and spatial scales. For example, the
Navier-Stokes partial differential equations describe the local in time and
space balance of forces in fluid flows. A static pressure difference leads to
material flow. However, despite the fact that any flow field is governed
instantaneously by these equations of motion, the equations themselves do not
directly describe fluid structures such as vortices, vortex pairs, vortex
streets, and vortex shedding, let alone turbulence \cite{Heis67a}. When
structures are generated at spatiotemporal scales far beyond those directly
specified by the equations of motion, we say that the patterns are
\emph{emergent}.

Two questions immediately come to the fore about emergent patterns. And, this
is where the subtlety arises \cite{Crut93g}. We see that something new has
emerged, but how do we objectively describe its structure and organization?
And, more prosaically, how do we discover patterns in the first place?

Refining the reconstruction methods developed to identify chaotic dynamics in
fluid turbulence \cite{Pack80,Take81}, computational mechanics
\cite{Crut88a,Crut12a} provided an answer that was as simple as it was
complete: a complex system's architecture lies in its causal states. A
\emph{causal state} is a set of histories, all of which lead to the same set of
futures. It's a simple dictum: Do not distinguish histories that lead to the
same predictions of the future.

The causal states and the transition dynamic over them give a canonical
representation---the \emph{\eM}. A system's \eM\ is its unique optimal
predictor of minimal size \cite{Crut88a,Crut98d,Shal98a}. The historical
information stored in the causal states of a process quantifies how structured
the process is. A process' \eM\ is its effective theory---its equations of
motions. One notable aspect of the \eM\ construction is that focusing on how to
optimally predict a process leads to a notion of structure. Predictability and
organization are inextricably intertwined.

With a system's \eM\ minimal representation in hand, the challenge of
quantifying emergent organization is solved. The answer lies in a complex system's \emph{intrinsic computation} \cite{Crut88a} which answers three simple questions:
\begin{enumerate}
\setlength{\topsep}{-2pt}
\setlength{\itemsep}{-3pt}
\setlength{\parsep}{-2pt}
\setlength{\labelwidth}{0pt}
\setlength{\itemindent}{-20pt}
\item How much of the past does a process store?
\item In what architecture is that information stored?
\item How is that stored information used to produce future behavior?
\end{enumerate}
The answers are direct: the stored information is that in the causal states;
the process' architecture is laid out explicitly by the \eM's states and
transitions; and the
production of information is the process' Shannon entropy rate.

At first blush it may not be apparent, but in this, computational mechanics
parallels basic physics. Physics tracks various kinds of energy and monitors
how they can be transformed into one another. Computational mechanics asks,
What kinds of information are in a system and how are they transformed into one
another? Although the \eM\ describes a mechanism that generates a system's
statistical properties, computational mechanics captures more than mere
generation. And, this is how it was named: I wished to emphasize that it was an
extension of statistical mechanics that went beyond analyzing a systems'
statistical properties to capturing its computation-theoretic properties---how
a system stores and processes information, how it intrinsically computes.

\vspace{-0.1in}
\section{Progress}
\vspace{-0.1in}

One might be concerned that this view of complex systems is either not well
grounded, on the one hand, or not practical, on the other. Over the last three
decades, however, computational mechanics led to a number of novel results from
theoretical physics and nonlinear mathematics that solidified its foundations
to applications that attest to its utility as a way to discover new science.
Discoveries from over the last decade or so give a sense of the power of the ideas
and methods, both their breadth and technical depth.

Recent theoretical physics and nonlinear mathematics contributions include the
following:
\begin{itemize}
\setlength{\topsep}{0pt}
\setlength{\itemsep}{-3pt}
\setlength{\parsep}{0pt}
\setlength{\labelwidth}{0pt}
\setlength{\itemindent}{-20pt}
\item Continuum and nonergodic processes 
	\cite{Marz14c,Marz17b,Marz14b, Marz15a, Marz14e,Marz14a,Crut15a,Trav11b,Crut15a};
\item Analytical complexity \cite{Riec16a,Crut13a,Riec13a,Riec17a,Riec14a,Riec14b};
\item Causal rate distortion theory \cite{Stil07a,Stil07b,Marz14f};
\item Synchronization and control \cite{Jame10a,Trav10a,Trav10b,Crut10a};
\item Enumerating memoryful processes \cite{John10a,Crut88a,Feld08a};
\item Crypticity and causal irreversibility
	\cite{Crut08a,Crut08b,Maho09b,Maho09a,Crut10d,Maho11a,Elli11a};
\item Bayesian structural inference \cite{Stre07b,Stre07a,Stre13a};
\item Input-output systems \cite{Barn13a};
\item Complexity of prediction versus generation \cite{Reub17a};
\item Predictive features and their dimensions \cite{Marz17a};
\item Sufficient statistics from effective channel states \cite{Jame17a};
\item Equivalence of history and generator \eMs\ \cite{Trav11a};
\item Informational anatomy \cite{Jame11a}; and
\item Automated pattern detection \cite{McTa04a}.
\end{itemize}

Recent applications of computational mechanics include the following:
\begin{itemize}
\setlength{\topsep}{0pt}
\setlength{\itemsep}{-3pt}
\setlength{\parsep}{0pt}
\setlength{\labelwidth}{0pt}
\setlength{\itemindent}{-20pt}
\item Complex materials \cite{Varn15a,Riec14b,Riec14a,Varn14a,Lei17a};
\item Stochastic thermodynamics
\cite{Marz17c,Agha16d,Riec16b,Boyd16c,Boyd14b,Boyd17a,Boyd15a,Boyd16e,Boyd16d};
\item Information fluctuations \cite{Agha16c,Crut16a};
\item Information creation, destruction, and storage \cite{Jame13a};
\item Spatiotemporal computational mechanics \cite{Rupe17a,Rupe17b};
\item Quantum mechanics \cite{Agha17a,Agha16b,Riec15b,Maho15a,Agha16a}; and
\item Evolution \cite{Crut10c,Crutchfield&Mitchell94a,Crut04a,Goer06a,Goer08a}.
\end{itemize}
Staying true to our present needs, this must leave out detailed mention of a
substantial body of computational mechanics research by others---a body that
ranges from quantum theory and experiment \cite{Gu12a,Pals15a,Thom17a} and
stochastic dynamics \cite{Ryab11a,Neru10a,Neru12a,Kell12a,Li08a,Witt97a} to
spatial \cite{Delg97a,Das96a,Hord99a,Gonc98a,Palm00a,Clar03a} and social
systems \cite{Darm13a}.

\vspace{-0.1in}
\section{History}
\vspace{-0.1in}

What's lost in listing results is the intellectual history of computational
mechanics. Where did the ideas come from? What is their historical context?
What problems drove their invention? Revisiting the conditions from which
computational mechanics emerged shows that aspects of this history resonate
with the science that followed.

My interests started as a fascination with mainframe computers in the 1960s
and with information theory in the 1970s. I worked for a number of years in
Silicon Valley, for IBM at what was to become its Almaden Research Center on
information storage technology---magnetic bubble devices---and at Xerox's Palo
Alto Research Center---which at the time was busily inventing our current
computing environment of packet-based networks (ethernet), internet protocols,
graphical user interfaces, file servers, bitmap displays, mice, and personal
workstations. An active member of the Homebrew Computer Club, I built
a series of microcomputers---4-bit, 8-bit, and eventually 16-bit machines.
There, I met many technology buffs, several who later become titans of
modern information technology. I suggested and then helped code up the first
cellular automaton simulator on a prototype 6502 (8-bit) microcomputer, which
would become the Apple I.

As a college student at the University of California, Santa Cruz (UCSC), I
learned about the mathematics of computers and communication theory directly
from the information theory pioneer David Huffman. Huffman, in particular, was
well known for his 1950s work on minimal machines---on what was called
\emph{machine synthesis}. His pioneering work was an integral part of his
discrete mathematics and information theory courses. Harry Huskey, one of the
engineers on the first US digital computers (ENIAC and EDVAC) also taught at
UCSC and I learned computer architecture from him.  In short, thinking about
computing and its physical substrates went hand in hand with my physics
training in statistical mechanics and mathematics training in dynamical systems
theory. This theme drove the bulk of my research on chaotic dynamics.  

With this background in mind, let me turn to address what were the immediate
concerns of nonlinear physics in the 1980s. As computers reduced in size and
cost, they became an increasingly accessible research tool. In the late 1970s
and early 1980s it was this revolution that led to the burgeoning field of
nonlinear dynamics. In contrast with abstract existence proofs, through
computer simulations we could simply look at and interact with the solutions of
complex nonlinear systems. In this way, the new tools revealed, what had been
relatively abstract mathematics through most of the $20^{\mathrm{th}}$ century,
a new universe of exquisitely complex, highly ramified structures and unpredictable behaviors.

Randomness emerged spontaneously, though paradoxically we knew (and had
programmed) the underlying equations of motion. This presented deep challenges.
What is randomness? Can we quantify it? Can we extract the underlying equations
of motion from observations? Soberingly, was each and every nonlinear system,
in the vast space of all systems, going to require its own ``theory''?  The
challenge, in essence, was to describe the qualitative properties of complex
systems without getting bogged down in irrelevant explicit detail and
microscopic analysis. How to see the structural forest for the chaotic trees?

In the 1970s a target problem to probe these questions was identified by the
nonlinear physics community---fluid turbulence---and a testable
hypothesis---the Ruelle-Takens conjecture that strange attractors were the
internal mechanism driving it \cite{Ruel71a}. This formalized an earlier
proposal---``deterministic nonperiodic flow''---by the meteorologist Lorenz
\cite{Lore63a}: nonlinear instability was responsible for the unpredictability
of weather and fluid turbulence generally.

There was a confounding problem, though. On the one hand, we had time series of
measurements of the fluid velocity at a point in a flow. On the other, we had
the abstract mathematics of strange attractors---complicated manifolds that
circumscribed a system's instability. How to connect them? This was solved by
the proposals to use the measured time series to ``reconstruct'' the system's
effective state space. This was the concept of extracting the attractor's
``geometry from a time series'' (1980-81) \cite{Pack80,Take81}. These
reconstruction methods created an effective state space in which to look at the
chaotic attractors and to quantitatively measure their degree of instability
(Kolmogorov-Sinai entropy and Lyapunov characteristic exponents) and their
attendant complicatedness (embedding and fractal dimensions). This was finally
verified experimentally in 1983 \cite{Bran83}, overthrowing the decades-old
Landau-Lifshitz multiple incommensurate-oscillator view of turbulence.

Reconstructing a chaotic attractor from a time series became a widely used
technique for identifying and quantifying deterministic chaotic behavior,
leading to the field of nonlinear time series modeling \cite{Casd91a}.

Reconstruction, however, fell short of concisely expressing a system's internal
structure. Could we extend reconstruction to extract the system's very
equations of motion? A substantial benefit would be a robust way to predict
chaotic behavior. The answer was provided in a method to reconstruct
``Equations of Motion from a Data Series'' \cite{Crut87a,Farm87}.

This worked quite well, when one happened to choose a mathematical
representation that matched the class of nonlinear dynamics generating the
behavior. But as Ref. \cite{Crut87a} demonstrated in 1987, if you did not have the
correct representational ``basis'' it not only failed miserably, it also did
not tell you how and where to look for a better basis. Thus, even this approach
to modeling complex systems had an inherent subjectivity in the choice of
representation. Structural complexity remained elusive.

How to remove this subjectivity? The answer was provided by pursuing a
metaphor to the classification scheme for automata developed in discrete
computation theory \cite{Chom56,Mins67,Hopc79}. There, the mathematics of
formal languages and automata had led in the 1950s and 1960s to a structural
hierarchy of representations that went from devices that used finite memory to
infinite memories organized in different architectures---tapes, stacks, queues,
counters, and the like.

Could we do this, not for discrete bit strings, but continuous chaotic
systems?  Answering this question led directly to computational mechanics as
laid out in 1989 by Ref. \cite{Crut88a}. The answer turned on a predictive
equivalence relation developed from the geometry-of-a-time-series concept of
reconstructed state \cite{Pack80} and adapted to an automata-theoretic setting.
The equivalence relation gave a new kind of state that was a distribution of
futures conditioned on past trajectories in the reconstructed state space.
These were the causal states and the resulting probabilistic automata were
\eMs. In this way, many of the notions of information processing and computing
could be applied to nonlinear physics.

\vspace{-0.1in}
\section{Misdirection}
\vspace{-0.1in}

The preceding history introduced the goals of computational mechanics, showed
its recent progress, and put its origins in the historical context of nonlinear
dynamics of complex systems, such as fluid turbulence. As we will now see, the
original history and recent progress form a necessary backdrop for some
distracting, but pressing business.

It is abundantly clear at this point that the preceding overview is \emph{not a
literature review} on intrinsic computation embedded in complex systems. Such a
review would be redundant since reviews and extensive bibliographies that cite
dozens of active researchers have been provided elsewhere and at semi-regular
intervals since Ref. \cite{Crut88a} (1989); see, e.g., Refs.
\cite{Crut92c,Crut98d,Crut01a,Shal98a,Crut12a}. Rather, the preceding is
provided as a narrative synopsis of its motivations, goals, and historical
setting. After three decades of extensive work by many researchers in
computational mechanics, why is this necessary? The reason is that critiques
appeared recently that concern computational mechanics publications from the
1980s and 1990s---that is, works that are two and three decades old. And so,
the early history and recent progress is a necessary backdrop.

The following addresses the issues raised and explains that, aside from several
interesting, detailed  mathematical issues, they are in large measure
misguided. They are based on arguments that selectively pick details, either
quoting them out of context or applying inappropriate contexts of
interpretation. As presented, they are obscured technically so that expertise
is required to evaluate the arguments. In other cases, the issues raised are
not criticisms at all---they are already well known. The following (A) reviews
the issues and offers a broad response that shows they are misguided at best
and (B) highlights the rhetorical style of argumentation, which shows that the
nontechnical (in some cases, ad hominem) arguments rely on fundamental errors
of understanding. After reviewing all of them carefully, we cannot find any
concern that would lead one to question the very solid and firm grounding of
computational mechanics.

\vspace{-0.1in}
\subsection{Technical Contentions}
\vspace{-0.1in}

As analysis tools, \eMs\ are defined and used in two different ways. In the
first they are defined via the predictive equivalence relation over sequences,
as already discussed and as will be detailed shortly; these are \emph{history
\eMs}. In the second, \eMs\ are defined as predictive generators of processes;
these are \emph{generator \eMs}. (Mathematically, they are unifilar
hidden Markov models with probabilistically distinct states that generate a given process.) The definitions are
complementary. In the first, one goes from a given process to its \eM; in the
second, one specifies an \eM\ to generate a given process. Importantly, the
definitions are equivalent and this requires a nontrivial proof \cite{Trav11a}.
The criticisms concern history \eMs and so we need focus only on them. The
computational mechanics of \eM\ generators is not at issue.

Reference \cite{Gras17a} raises technical concerns regarding statistical estimation of finite-state and probabilistic finite state machines, as discussed in several-decades-old computational mechanics publications; principally two from 1989 and 1990: Refs. \cite{Crut88a} and \cite{Crut89e}, respectively.

The simplest response is that almost all of the concerns have been superseded
by modern computational mechanics: mixed-state spectral decomposition
\cite{Crut13a,Riec16a} and Bayesian structural inference and \eM enumeration
methods \cite{Stre13a,John10a}. The view from the present is that the
issues are moot.

That said, when taken at face value, the bulk of the issues arise from
technical misinterpretations. Largely, these stem from a failure to take into
account that computational mechanics introduced and regularly uses a host of
different equivalence relations to identify related, but different kinds of
state. Ignoring this causes confusion. Specifically, it leads to Ref.
\cite{Gras17a}'s misinterpretations of covers and partitions of sequence space,
transient versus recurrent causal states, the vanishing measure of
nonsynchronizing sequences, and an \eM's start state. It also leads to a second
confusion over various machine reconstruction methods. Let's take these two
kinds of misunderstanding in turn.

\paragraph*{Effective states and equivalence relations}
One of computational mechanics' primary starting points is to identify a
stochastic process' effective states as those determined by an equivalence
relation. Said most simply, group pasts that lead to the same distribution of
futures. Colloquially: do not make distinctions that do not help in
prediction. The equivalence relation $\sim$ connects two pasts $\ms{-K}{0} =
\meassymbol_{-K} \ldots \meassymbol_{-1}$ and
$\ms{-K^\prime}{0} = \meassymbol_{-K^\prime} \ldots \meassymbol_{-1}$, if the
future $\MS{0}{L} = \MeasSymbol_0 \ldots \MeasSymbol_{L-1}$ after having seen each looks the same:
\begin{align*}
\ms{-K}{0} \sim \ms{-K^\prime}{0} \Leftrightarrow
  \Pr(\MS{0}{L}|\ms{-K}{0}) = \Pr(\MS{0}{L}|\ms{-K^\prime}{0})
  ~.
\end{align*}
Taking finite or infinite pasts and futures and those of equal or unequal
lengths defines a family of equivalence relations and so of different kinds of
causal state.

``Inferring Statistical Complexity'' (1989) focused on determining a process'
long-term memory and so used $K, K^\prime \to \infty$ and $L \to \infty$ \cite{Crut88a}.
That is, it worked with a process' \emph{recurrent causal states}, defining the
process' \emph{statistical complexity} as the amount of information they
store. This and later works also used finite pasts ($K, K^\prime \in \{0, 1,
2, 3, \ldots\}$) and infinite futures ($L \to \infty$) to define causal states
more broadly. This introduced the notion of \emph{transient causal states}.
In turn, they suggested the more general notion of \emph{mixed states}
that monitor how an observer comes to know a process' effective states---how
the observer \emph{synchronizes} to a process.
And, finally, in this regime one has the
\emph{subtree reconstruction method} that merges candidate states with
different-length pasts. The mixed
states are critical to obtaining closed-form expressions for a process'
information measures \cite{Crut13a,Riec13a,Riec17a}. This setting also
introduces the notion of an \eM's \emph{start state}---the effective state the
process is in, having a correct model in hand, but having made \emph{no
measurements}: $K, K^\prime = 0$. Similarly, later works used infinite pasts
and finite-length futures. Finally, using pasts and futures of equal length,
but increasing them incrementally from zero leads to the class of
\emph{causal-state splitting reconstruction} methods \cite{Shal02a}.

Why all these alternatives? The answer is simple: each equivalence relation in
the family poses a different question to which the resulting set of states is
the answer or, at least, is an aid in answering. For example and somewhat
surprisingly, Upper showed that even with infinite pasts and futures and the
induced recurrent causal states, there are \emph{elusive} and
\emph{unreachable} states that are never observed \cite{Uppe97a}. More to the
point, defining other kinds of state has been helpful in other ways, too. For
example, to define and then calculate a process' \emph{Markov} and
\emph{cryptic orders} requires a different kind of transient state
\cite{Jame10a}. Analogously, very general convergence properties of stochastic
processes are proved by constructing the states of a process' \emph{possibility
machine} \cite{Trav10a,Trav10b,Trav14a}.

With this flexibility in defining states, the mathematical foundations of
computational mechanics give a broad set of analytical tools that tell one how a
given process is organized, how it generates and transforms its information.
Insisting on and using only one definition of causal state gives a greatly
impoverished view of the structure of stochastic processes. Each kind is an
answer to a different question. Apparently, this richness and flexibility is a
source of confusion. No surprise, therefore, that if a question of interest is
misunderstood, then a given representation may appear wrong, when it is
in fact correct for the task at hand.

\paragraph*{Reconstruction methods}

Reference \cite{Gras17a} is unequivocal in its interpretation of machine
reconstruction. It turns out there is little need to go into a detailed
rebuttal of its statements, as they arise from a kind of misinterpretation
similar to the misinterpretations discussed above. In short, Ref.
\cite{Gras17a} confuses a set of related, but distinct machine reconstruction
methods.

For example, sometimes one is interested in a representation of the state
machine that simply describes a process' set of allowed realizations; that is,
we are not interested in their probabilities, only which strings occur and
which do not. This is the class of \emph{topological machine reconstruction}
methods; the origins of which go back to the earliest days of the theory of
computation---to David Huffman's work. One can also, as a quick approximation,
take a topologically reconstructed machine and have it read over a process'
sequence data and accumulate transition and state probabilities. This is a
mixture of topological reconstruction and empirical estimation. And, finally,
one can directly estimate fully probabilistic \eMs via algorithms that
implement the equivalence relation of interest.

One can then use this range of reconstruction methods---topological,
topological plus empirical, and probabilistic---with one or the other of
the above equivalence relations.

It is important to point out that these statistical methods all have their
weaknesses. That is, for a given reconstruction algorithm implementation, one
can design a process sample for which the implementation will behave
misleadingly. For example, it has been known for some time that causal-state
splitting reconstruction methods \cite{Shal02a} often give machines with a
diverging set of states, if one presents them with increasingly more data. This
occurs due to its ``determinization'' step, which has an exponential state-set
blow-up when converting an intermediate, approximate nondeterministic
presentation to a deterministic (or unifilar) one. Analogously, the subtree
reconstruction method suffers from ``dangling states'' in which inadequate data
leads to improperly estimated future conditional distributions from which there
is no consistent transition. This is not surprising in the least. Many arenas
of statistical inference are familiar with such problems, especially when
tasked to do out-of-class modeling. The theoretical sleight of hand one finds
in mathematical statistics is to \emph{assume} data samples come from a known
model class. For those interested in pattern discovery, this begs the question
of what are patterns in the first place.

Now, many such problems can be overcome in a theoretical or computational
research setting by presenting the algorithms with a sufficient amount data.
However, in a truly empirical setting with finite data, one must take care in
their use.

To address the truly empirical setting, these problems led us to introduce
\emph{Bayesian Structure Inference} for \eMs \cite{Stre13a}. It relies on an
exact enumeration of a set of candidate \eMs and related models
\cite{John10a}. It does not suffer from the above estimation problems in that
it does not directly convert data to states and transitions as the above
reconstruction algorithms do. Rather, it uses well-defined candidate models
(\eMs) to estimate the probability that each produced the given data. It works
well and is robust, even for very small data sets. That is, it is data
parsimonious and relatively computationally efficient. And, if one has extra
knowledge (from theoretical or symmetry considerations) one needs only use a
set of candidate models consistent with that knowledge. In many settings, this
leads to markedly increased computational efficiency.

To close this section, it is clear that one could spend an inordinate amount of
time arguing which combination of the above equivalence relations and
reconstruction methods is ``correct'' and which is ``incorrect''. This strikes
me as unnecessarily narrow. The options form a toolset and those methods that
produce consistent results, strengthened by testing against known cases, yield
important process properties. Practically, I recommend Bayesian Structural
Inference \cite{Stre13a}. If I know a source will have low entropy rate and I
want to see if it is structurally complex, though, I use probabilistic subtree
reconstruction. I avoid causal-state splitting reconstruction.

\vspace{-0.1in}
\subsection{Rhetorical Diversions}
\vspace{-0.1in}

The preceding text offers a concise rebuttal to Ref. \cite{Gras17a}'s 
claims by identifying their common flaws. The latter's technical discussion, though, is
embedded in a misleading rhetorical style. The import of this misdirection may
be conveyed by analyzing two less technical points that are also presented with
distracting emotion.

The first is a misreading of the 1989 computational mechanics publication,
``Inferring Statistical Complexity''. The claim is that the title is
grossly misleading since the article is not about statistical inference. This
is an oddly anachronistic view of work published $30$ years ago, which
seems to require looking through the lens of our present Big Data era and the
current language of machine learning.

Read dispassionately, the title does allude to ``inferring'', which the
dictionary says is ``deducing or concluding (information) from evidence and
reasoning rather than from explicit statements''. And that, indeed, is how the
article approaches statistical complexity---discovering patterns of intrinsic
computation via the causal equivalence relation. It not only defines
statistical complexity, but also introduces the mathematics to extract it. Yes,
the article is not statistical inference. The topic of statistical inference as
it is understood today was  addressed in a number of later works; the most
recent of which was mentioned above---``Bayesian Structural Inference for
Hidden Processes'' \cite{Stre13a}. In short, the criticism is as specious as
the rhetoric is distracting: the claim attributes anachronistic and inaccurate
meanings to the article.

The second nontechnical issue is developed following a similar strategy, and it
also reveals a deep misunderstanding. Packard and I had studied
the convergence properties of Shannon's entropy rate
\cite{Crut82b,Crut82c,Crut83a} and along with Rob Shaw \cite{Shaw84} had
realized there was an important complexity measure---the past-future mutual
information or \emph{excess entropy}---that not only controlled convergence,
but was on its own a global measure of process correlation. As those articles
and Packard's 1982 PhD dissertation \cite{Pack82} point out this quantity was
already used to classify processes in ergodic theory \cite{Junc79}.

Given that excess entropy's basic properties and alternative definitions had
been explored by then, Packard and I moved on to develop a more detailed
scaling theory for entropy convergence, as one of the articles noted in its title
``Noise Scaling of Symbolic Dynamics Entropies''. In this we defined the
\emph{normalized excess entropy}, which was normalized to its exact
infinite-history, zero-noise value. This followed standard methods in phase
transition theory to use ``reduced'' parameters. (A familiar example is the
\emph{reduced temperature} $\widehat{t} = (T - T_c) / T_c$ normalized to vanish
at the critical temperature $T_c$ at which the phase transition of interest
occurs.)

The complaint is that this definition is intentionally misleading since it is
not the excess entropy. Indeed, it is not. The normalized excess entropy is a
proxy for a single term in the excess entropy. And, the article is absolutely
clear about its focus on scaling and the tools it employs. Once one does have a
theory of how entropy convergence scales, in particular the convergence rate,
then it is easy to back-out the excess entropy. A simple formula expresses the
excess entropy in terms of that rate and the single-symbol entropy.

So, this too is a toothless criticism, but it exemplifies the emotion and rhetorical style employed throughout Ref. \cite{Gras17a}. The nontechnical and ad hominem criticisms intertwined with the technical faults are evidence of the consistent projection of irrelevant meanings onto the material. Once such an intellectually unproductive strategy is revealed, further rebuttal is unnecessary.

\section{Final Remarks}

To summarize, computational mechanics rests on firm foundations---a solidity
that led to many results over the last three decades, ranging from theoretical
physics and nonlinear mathematics to diverse applications. It is a direct
intellectual descendant of many researchers' efforts, including my own, in the
1970s and early 1980s to describe the complex behaviors found in fluid
turbulence.

Reference \cite{Gras17a}'s technical claims arise from a misunderstanding of
computational mechanics' goals, methods, successes, and history. Its rhetoric
reveals a strategy of quoting out of context and reinterpreting decades-old
work either without benefit of modern results or projecting arbitrary
assumptions onto the early work. Any dogmatic conclusions on what is
``correct'' that follow from such a strategy are flawed. Moreover, Ref.
\cite{Gras17a}'s claims to precedence are based on false memories, are
unsubstantiated, and, in light of the history of events, are unsubstantiatable.

Current work simply eclipses the questions raised in distant retrospect, rendering the criticisms moot. Time passes. We should let it move on.

Over the years, computational mechanics has been broadly extended and applied,
far beyond its initial conception $30$ years ago. That said, its hope to lay
the foundations of
a fully automated ``artificial science'' \cite{Crut88a}---in which
theories are built automatically from raw data---remains a challenge. Though
the benefits are tantalizing, it was and remains an ambitious goal.

\acknowledgments

I thank many colleagues who, over the years, improved, corrected, and extended
computational mechanics using constructive debate and civil correspondence and
who made the effort to understand its motivations, its contributions, and its
limitations. Most immediately, I am grateful to Cina Aghamohammadi, Korana
Burke, Chris Ellison, Jeff Emenheiser, Dave Feldman, David Gier, Mile Gu,
Martin Hilbert, Ryan James, John Mahoney, Sarah Marzen, Norman Packard, Geoff
Pryde, Paul Riechers, Dawn Sumner, Susanne Still, Meredith Tromble, Dowman
Varn, Greg Wimsatt, Howard Wiseman, and Karl Young for helpful comments. I
thank the Santa Fe Institute for its hospitality during visits, where I have
been a faculty member for three decades. This material is based upon work
supported by, or in part by, John Templeton Foundation grant 52095,
Foundational Questions Institute grant FQXi-RFP-1609, the U.S. Army Research
Laboratory and the U. S. Army Research Office under contracts W911NF-13-1-0390,
W911NF-13-1-0340, and W911NF-12-1-0288.

\end{document}